%% file: apssamp.tex
\documentclass[%
 reprint,
 amsmath,amssymb,
 aps,
]{revtex4-2}

\usepackage{graphicx}
\usepackage{dcolumn}
\usepackage{bm}
\usepackage{hyperref}

\usepackage{xcolor}

\begin{document}

\preprint{APS/123-QED}

\title{Flocking with random non-reciprocal interactions
}

\author{Jiwon Choi$^1$, Jae Dong Noh$^2$, Heiko Rieger$^1$}
\affiliation{
$^1$ Department of Physics \& Center for Biophysics, Saarland University, Campus E2 6, 66123 Saarbrücken, Germany\\
$^2$ Department of Physics, University of Seoul,
Seoul, 02504, Korea}


\date{\today}

\begin{abstract}
Flocking is ubiquitous in nature and emerges due to short or long range 
alignment interactions among self-propelled agents. 
Two unfriendly species that anti-align or even interact non-reciprocally show more complex collective phenomena, ranging from parallel and anti-parallel flocking over run-and chase behavior to chiral phases.
Whether flocking or any of these collective phenomena can survive in the presence of a large number of species with random non-reciprocal interactions remained elusive so far. As a first step here the extreme case of a Vicsek-like model with fully random non-reciprocal interactions between the individual particles is considered. As soon as the 
alignment bias is of the same order as the random interactions 
the ordered flocking phase occurs, but deep within this phase, the random non-reciprocal interactions can still support global chiral and oscillating
states in which the collective movement direction rotates or oscillates slowly. For short-range interactions,
moreover, even without alignment bias 
self-organized cliques emerge, in which medium size clusters of particles that have predominantly aligning interactions meet accidentally and stay together for macroscopic times. These results may serve as a starting point for the study of multi-species flocking models with non-random but complex non-reciprocal inter-species interactions.

\end{abstract}

\maketitle



\section{Introduction}
Active matter consists of particles that consume energy to use it 
as a fuel for movement, force generation, deformation or proliferation
\cite{AM-Review-Marchetti,AM-Review-Ramaswami,AM-Review-Bechinger,AM-Review-Rieger} and gives rise to collective phenomena that are 
absent in equilibrium systems, like flocking~\cite{Vicsek1995Novel},
motility induced phase separation \cite{MIPS-Review},
spontaneously flowing matter~\cite{Bricard2013emergence,Schaller2010polar} or living crystals~\cite{Theurkauff2012dynamic,Palacci2013living}.
Recently active matter with non-reciprocal interactions (NRI)
has attracted a lot of attention, for which 
paradigmatic examples are the predator-prey relationship, 
visual perception~\cite{Lavergne2019group,Barberis2016large}
or flocking of unfriendly species~\cite{Fruchart2021nonreciprocal,Kreienkamp2024nonreciprocal,Mangeat2025emergent,Avni2025NRAIM,Martin2023transition}.
The hallmark of NRI is the emergence of a broad range of 
fascinating phenomena ranging from the spontaneous emergence of 
travelling waves \cite{You2020nonreciprocity,Brauns2024nonreciprocal}
and chiral states \cite{Fruchart2021nonreciprocal} 
to odd viscosity \cite{OddViscosity-Vitelli}, elasticity \cite{OddElasticity-Vitelli} and diffusivity \cite{OddDiffusivity-Hargus,OddDiffusivity-Kalz}.

In systems consisting of particles with only two-body interactions 
non-reciprocity manifests itself by the violation of Newton's law
"{\it actio=reactio}" such that the force or influence exerted by particle $j$ on particle $i$, $F_{ij}$ is {\it not} equal to $-F_{ji}$.
In complex systems with many species of particles NRI can be encoded in a non-symmetric interaction matrix $J_{ij}$, as for instance
by non-symmetric reaction rates for chemical mixtures or 
predator-prey systems. For complex systems consisting of a large number 
of species this interaction matrix is frequently assumed to be random as in the context of neural networks 
\cite{Sompolinsky1986temporal,crisanti1987dynamics,Rieger1988,Sompolinsky1988chaos},
ecological communities \cite{Rieger1989solvable,Altieri2021properties,Ros2023generalized,Lorenzana2024},
the immune system~\cite{Parisi1990simple}, 
coupled oscillators~\cite{Stiller1998dynamics,Prueser2024nature,Prueser2024role,Hanai2024nonreciprocal} and multi-species suspensions~\cite{Parkavousi2025enhanced}.
Random-field type disorder, meaning heterogeneous environments instead of interactions, has been studied for the Vicsek model \cite{Duan2021breakdown,Chepizhko2013optimal,Chardac2021emergence,Morin2016distortion,Codina2022small} and scalar active matter~\cite{Ro2021disorder,BenDor2022disorder},
but the impact of random non-reciprocal (NR) alignment interactions on 
flocking systems remains elusive.

Large ensembles of self-propelled particles 
may consist of many species with varying inter-species interactions, each of which can be aligning for friendly species or anti-aligning 
for unfriendly species \cite{TSVM-Noh}. In the simplest case of only two species NRI gives rise to run-and chase states, a travelling wave in which one species follows the other, and chiral states, in which all particles perform a coherent rotation \cite{Martin2023transition,Kreienkamp2024nonreciprocal,Fruchart2021nonreciprocal,Mangeat2025emergent}, in addition to the still possible flocking state, in which all particles move collectively in the same direction. Whether these collective phenomena can survive in the presence of NRI in a multi-species system is the question that we will address in this paper. Under which circumstances can flocking still occur or are chiral states still possible? If global order is destroyed, could local order in the form of micro-flocks or cliques occur? As a first step to shed light into these problems 
we consider here a Vicsek-like model with random NRI and tunable alignment bias between the individual particles. This setting includes two different types of
scenarios:  1) a group of flocking animals of the same species is comprised of individuals with a common alignment bias but assigns 
NR corrections to the directional information received from other individuals, corresponding to weak NR randomness, and 2) a multi-species ensemble of self-propelled particles, in which each particle belongs to a different species and has its own NR alignment/ant-alignment rules with each of the other particles, corresponding to strong NR randomness.
We will show that both scenarios bear the potential for collective  phenomena absent in the conventional one- or two-species flocking models.

The paper is organized as follows: in section II we define the model
for which we first analyze the infinite range limit in section III.
Then, in section IV we discuss finite range interactions and discuss our results in section V.

\section{Model}
We consider $N$ self-propelled particles or agents $i$, characterized by
a position ${\bf r}_i(t)\in L^d$ in a $d$-dimensional cube of lateral size $L$ ($d=2$ in this paper, p.b.c.) and a movement direction $\theta_i(t)\in[0,2\pi]$. 
Their mutual tendency to move in the same direction as friendly neighbors
(alignment) or to move in the opposite direction as unfriendly neighbors (anti-alignment) is described by a (noise-less) Vicsek-like model \cite{Vicsek1995Novel,Gregoire2004onset}
defined by the equations of motion
\begin{align}
    \dot{\mathbf r}_i(t) &= v_0\,(\cos(\theta_i(t)),\sin(\theta_i(t))) \label{eq1}\\
    \dot\theta_i(t) &= -\sum_{j\in\mathcal N_i} 
    \left(
    \frac{J_{ij}}{\vert\mathcal N_i\vert^{\alpha}}
    +\frac{J_0}{\vert\mathcal N_i\vert}\right)
    \sin(\theta_i(t)-\theta_j(t))
    \label{eq2}
\end{align}
The alignment/anti-alignment interactions consist of an alignment bias 
$J_0\ge$ and of a random 
part $J_{ij}$ which can be aligning $J_{ij}>0$ or anti-aligning $J_{ij}<0$. 
The $J_{ij}$ are Gaussian random variables with
$\overline{J_{ij}} = 0$, $\overline{J_{ij}^2} = J^2$, and 
$\overline{J_{ij}J_{ji}} = \lambda J^2$, where $\lambda\in[-1,1]$
controls the degree of non-reciprocity. 
Concretely, each interaction pair ($J_{ij}$, $J_{ji}$) is drawn from a bivariate Gaussian distribution
\begin{equation}
    P(J_{ij},J_{ji}) = \frac1{2\pi J^2\sqrt{1-\lambda^2}}
    \exp\left[-\frac{J_{ij}^2-2\lambda J_{ij}J_{ji}+J_{ji}^2}{2J^2(1-\lambda^2)}\right]
\end{equation}
In the limit $\lambda=\pm1$, $P(J_{ij},J_{ji})$        
becomes singular: for $\lambda=1$, the interaction pair is strictly reciprocal($J_{ij}=J_{ji}$), while for $\lambda=-1$ it is antisymmetric($J_{ij}=-J_{ji}$).
We set $J=1$ and measure $J_0$ and time $t$ in units of $J$.
$v_0$ denotes the self-propulsion speed.


$\mathcal N_i$ is the "neighborhood of particle $i$, comprising 
all particles $j$ in a distance smaller than $R$, and 
$\vert\mathcal N_i\vert$ is the number of particles in this neighborhood.
In the case of infinite range interactions ($R\to\infty$)
$\vert\mathcal N_i\vert=N$.
The exponent $\alpha\in\{1/2,1\}$ denotes different scalings of the interaction strengths: $\alpha=1/2$ has to be chosen in order to obtain a non-trivial mean-field limit $N,R\to\infty$, $\alpha=1$ is legitimate for any finite $N$ and weights bias and randomness similarly.
Coherent motion of a flock is indicated by the order parameter
$\vec{m}(t)=1/N\,\sum_i (\cos\theta_i(t),\sin\theta_i(t))=
m(t)\cdot(\cos\psi(t),\sin\psi(t))$,
and flocking occurs when $m(t)>0$.

\begin{figure}
    \includegraphics[width=\linewidth]{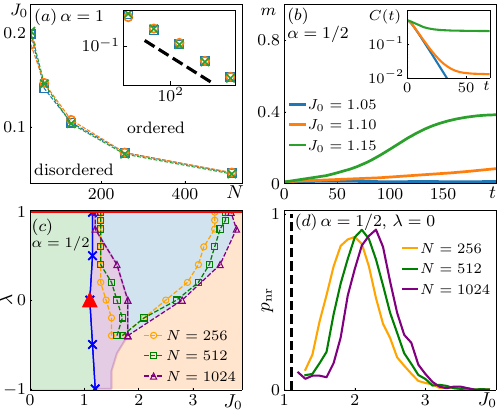}
    \caption{\label{fig:fig1}
        (a) Infinite range phase diagram for $\alpha=1$ and 
        $\lambda=-0.5$(circle), $0$(square), and $0.5$(cross).
        Inset: Log-log plot of the phase diagram. Black dashed line indicates $J_0\sim N^{-1/2}$.
        (b) Order parameter $m(t)$ (\ref{eq3}) for different $J_0$ at $\lambda=0$ obtained by solving the DMFT eq.(\ref{eq3}).
        The critical point is $J_{0,c}\approx1.1 $.
        Inset: Correleation function $C(t)$ for the same values of $J_0$ as in main figure.
        (c) $J_0$-$\lambda$ phase diagram for $\alpha=1/2$.
        Green: disordered phase, purple: flocking with relaxational dynamics(c.f. Fig.~\ref{fig:fig2}(a)); blue: flocking region with non-relaxational dynamics (c.f. Fig.~\ref{fig:fig2}(b-d)), c.f. (d). Note that this region shrinks with increasing $N$ (see text). Orange: static flocking. The blue line indicates the disorder-flocking phase boundary~\cite{supp}, the red line represents the spin glass region at $\lambda=1$.
        The red triangle locates the critical point from DMFT at 
        $\lambda=0$.
        (d) Fraction of disorder-realizations displaying non-relaxational dynamics $p_{\text{nr}}$ as a function of $J_0$ (at $\lambda=0$). The black dashed line indicates the critical point from DMFT. $p_{\text{nr}}>0.15$ defines the blue region in (d).
    }
\end{figure}

\begin{figure*}
    \includegraphics[width=\linewidth]{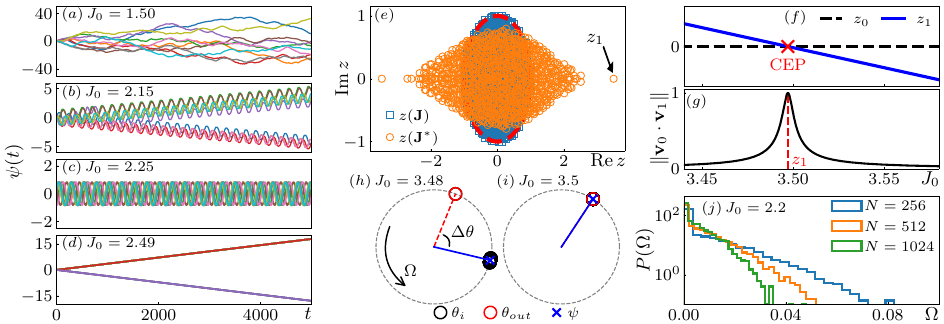}
    \caption{\label{fig:fig2}
        Emergence of non-relaxational dynamics for finite $N$ 
        with infinite-range interactions.
        (a-d) Time evolution of magnetization phase $\psi(t)$ for a single realization of interaction matrix $[\mathbf J]_{ij} = J_{ij}/\sqrt N$ with $N=1024$ and $\lambda=0$:
        (a) flocking with diffusive $\psi(t)$, (b) vibrating oscillation, (c) oscillation, and (d) rotation(see 'movie 2').
        Each line in (a-d) represents trajectories of $\psi(t)$ starting from different initial conditions.
        (e) Eigenvalue spectrum of the same interaction matrix $\mathbf J$ as in (a-d) and the corresponding linearized matrix $\mathbf J^\ast$, for $J_0=0$.
        The red dashed line indicates the circular law for $\lambda=0$~\cite{Sommers1988spectrum} and $z_1$ depicts the leading eigenvalue of $\mathbf J^\ast$, which locates the critical exceptional point(CEP): 
        (f) Decreasing $J_0>z_1$ shifts the spectrum to the right, and $z_1$ hits zero at the CEP ($z_1\approx3.497$) and (g) overlap between eigenvector $\mathbf v_1$ associated with $z_1$ and the Goldstone mode $\mathbf v_0$ reaches 1, indicating two eigenvectors coalesce.
        (h,i) Steady-state angle configuration near the CEP.
        (h) For $J_0=3.48<z_1$, the outlier component $\theta_{\text{out}}$ is seperated from the others with constant angle difference $\Delta\theta = \theta_{\text{out}}-\psi$ and drives global chiral motion with constant angular velocity $\Omega$.
        (i) For $J_0=3.50>z_1$, all angle are perfectly aligned.
        (j) Probability distribution of the time averaged angular velocity $P(\Omega)$ in the steady state for fixed $J_0=2.2$ and varying $N$.
    }
\end{figure*}

\section{Infinite-range limit}
{\bf Phase behaviors.}
For $R\to\infty$, the particle-particle interactions are independent of the particle positions, ${\bf r}_i$, and the angles $\theta_i$ 
evolve according to eq. (\ref{eq2}) with $\vert\mathcal N_i\vert=N$.
Flocking occurs when $J_0$ is larger than a critical alignment bias,
which for $\alpha=1$
decreases with $N$ as $J_{0,c}/\sqrt{N}$(Fig.~\ref{fig:fig1}(a)), 
with $J_{0,c}=1.1$, implying that flocking emerges for any $J_0>0$ in the limit $N\to\infty$.

In contrast, for $\alpha=1/2$, eq.~(\ref{eq2}) reduces in the limit $N\to\infty$ to a self-consistent stochastic single particle dynamics or dynamical mean-field theory (DMFT)~\cite{crisanti1987dynamics,Rieger1988,Stiller1998dynamics,Prueser2024role,Guilia2024nonreciprocal}.
For $\lambda=0$, this DMFT is given by
\begin{equation}
    \dot\theta(t) = (\varphi_y(t)+J_0 m_y(t))\cos\theta(t) - (\varphi_x(t)+J_0 m_x(t))\sin\theta(t)
    \label{eq3}
\end{equation}
where $\varphi_a(t)$ for $a=x,y$ is Gaussian colored noise with zero mean and correlations $\langle \varphi_a(t)\varphi_a(t')\rangle = C_{ab}(t,t')$ which are self-consistently given by the dynamics of $\theta$ via 
$C_{xx}(t,t')=\langle\cos\theta(t)\cos\theta(t')\rangle_\varphi$, 
$C_{yy}(t,t')=\langle\sin\theta(t)\sin\theta(t')\rangle_\varphi$, etc.
and the magnetization $(m_x(t),m_y(t)) = (\langle\cos\theta(t)\rangle_\varphi,\langle\sin\theta(t)\rangle_\varphi)$.
By solving eq.~\eqref{eq3} numerically~\cite{Roy2019numerical}, 
we determine for $\lambda=0$ the time dependence of the order 
parameter $m(t)$ (Fig.~\ref{fig:fig1}(b))
and the stationary auto-correlation function 
$C(t) = \lim_{t_0\rightarrow\infty} C(t+t_0,t_0)$ 
(Fig.~\ref{fig:fig1}(b), inset), which approach exponentially 
fast either zero in the disordered phase or $m_s>0$ and 
$m_s^2$, respectively, in the flocking phase.
These results also confirm the flocking transition to be at $J_{0,c}=1.1$ for $\lambda=0$.
The complete $J_0$-$\lambda$ phase diagram obtained by numerically integrating (\ref{eq1}) and performing finite-size scaling~\cite{supp} is shown in Fig.~\ref{fig:fig1}(c)).

%



\begin{figure}[h!]
    \includegraphics[width=\linewidth]{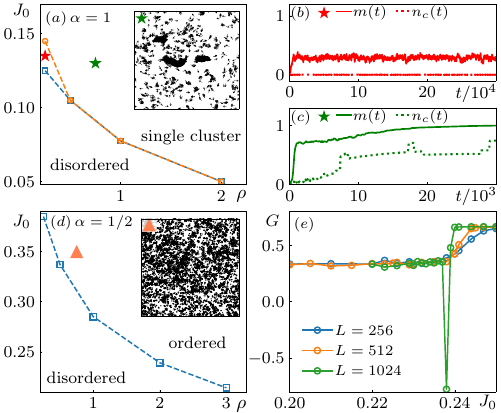}
    \caption{\label{fig:fig3}
        Finite range interactions ($R=1$, $\lambda=0$):
        (a) Phase diagram for $\alpha=1$.
        Region between the disordered and the single-cluster phase indicates long range order $m>0$ without condensation.
        Inset: Configuration snapshot ($L=1024$) of growing clusters in the single cluster region at the point marked with green star in (a).
        (b,c) Time evolution of magnetization $m(t)$ and the relative size of the largest cluster $n_c(t)$ at the red and green point marked in (a).
        (d) As in (a) for $\alpha=1/2$.
        Inset: Configuraton snapshot in the ordered region at the 
        orange triangle in (c).
        (e) Binder cumulant $G=1-\langle m^4\rangle/(3\langle m^2\rangle^2)$ at the transition ($\rho=2$).
    }
\end{figure}

\begin{figure*}
    \includegraphics[width=\linewidth]{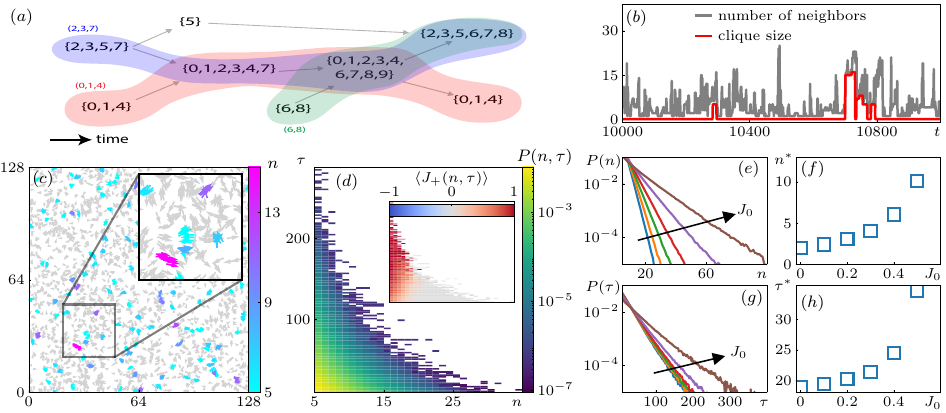}
    \caption{\label{fig:fig4}
        Clique formation for finite range interactions 
        ($R=1$, $\lambda=0$, $\alpha=1/2$, $\rho=0.25$).
        (a) Schematic diagram of the clique detecting algorithm~\cite{Buchin2015trajectory}. Each set $\{\cdots\}$ represents connected components and arrow depicts particle flows between connected components in different time.
        The colored regions indicate three potential cliques:
        $\{2,3,7\}$(blue), $\{0,1,4\}$(red), and $\{6,8\}$(green).
        (b) Time evolution of the number of interacting neighbors of a specific particle and the size of the largest clique containing it. 
        (c) Snapshot of a particle configuration for $J_0=0$.
        Cliques are represented with colors according to its size.
        Inset: Zoomed-in snapshot of the region indicated by the gray square.
        (d) $P(n,\tau)$ at $J_0=0$ averaged over many disorder realization.
        Inset: $\langle J_+(n,\tau)\rangle$ at the same parameters as (d).
        (e-f) $P(n)$ for different $J_0$ values and its characteristic cluster size $n^\ast$.
        (g-h) $P(\tau)$ for different $J_0$ values and its characteristic lifetime $\tau^\ast$.
    }
\end{figure*}

{\bf Non-relaxational dynamics for finite $N$.}
In contrast to the purely relaxational behavior of the DMFT
($N\to\infty$) of uncorrelated randomness
($\lambda=0$), the dynamics of a finite 
system ($N<\infty$) is much more interesting and 
can exhibit various non-relaxational dynamics such as oscillations, rotations and vibrating oscillations as shown in (Fig.~\ref{fig:fig2}(b-d) and 'movie2'). 
These attractors emerge in the whole blue region 
of Fig.~\ref{fig:fig1}(c) within the flocking phase, and persist 
up to, but not including, $\lambda=1$.
At $\lambda=1$ eq.(\ref{eq2}) predicts a zero-temperature spin glass
phase in which the angles $\theta_i$ relax into a local energy minimum of the underlying XY-Hamiltonian and freeze.
We analyze the non-relaxational behavior for finite $N$
by linearizing the equation of motion by $\theta_i = \psi + \delta_i$ for $J_0\gg J_{0.c}$.
The linearized equation of motion is then
\begin{equation}
    \dot{\vec\delta} = \mathbf J^\ast \vec\delta\;,
\end{equation}
where $[\mathbf J^\ast]_{ij} = J_{ij}/\sqrt N + J_0/N$ for $i\neq j$ and $[\mathbf J^\ast]_{ii} = -\sum_{j\neq i}[\mathbf J^\ast]_{ij}$.
While the eigenvalue spectrum of the original interaction matrix $\mathbf J$ follows the circular law~\cite{Sommers1988spectrum}, that of the linearized matrix $\mathbf J^\ast$ at $J_0=0$ is deformed due to rotational symmetry $\sum_j [\mathbf J^\ast]_{ij} = 0$ (Fig.~\ref{fig:fig2}(e)).
The ferromagnetic bias $J_0>0$ only shifts the entire eigenvalue spectrum of $\mathbf J^\ast$ by $-J_0$ along the $\text{Re}\,z$ axis, except for the Goldstone mode $\mathbf v_0=N^{-1/2}(1,\cdots,1)^T$, whose eigenvalue remains at zero\cite{supp}.
Therefore, for $J_0>z_1$, where $z_1$ denotes the leading eigenvalue of $\mathbf J^\ast$, all eigenmodes decay to zero, and the system settles into a static ordered phase with perfect alignment $\theta_i=\psi$(Fig.~\ref{fig:fig2}(i)).

As $J_0$ decreases and approaches $z_1$, the leading eigenvalue $z_1$ hits the Goldstone mode at zero(Fig.~\ref{fig:fig2}(f)), and the associated eigenvector $\mathbf v_1$ coalesces with the $\mathbf v_0$(Fig.~\ref{fig:fig2}(g)).
This is the hallmark of a critical exceptional point (CEP) \cite{Hanai2020critical,Zelle2024universal,Suchanek2023time}
beyond which chiral motion emerges.
At the CEP, all angles still stay close and begin to rotate collectively.
As $J_0$ decreases further, one "outlier" $\theta_{\text{out}}$ 
gradually deviates from the the other angles but continues to
rotate with them (Fig.~\ref{fig:fig2}(h)).
As soon as the angle-difference reaches $\pi$, the chiral motion 
stops and the static ordered state is restored\cite{supp} 
and remains stable until a second instability arises.
The nature of the second instability can be a Hopf bifurcation or again 
a critical exceptional point, depending on eigenvalue spectrum.
In the bulk spectrum, where multiple instabilities contribute, we observe non-relaxational behaviors shown in Fig.~\ref{fig:fig2}(b-d) at sufficiently high $J_0$(Fig.~\ref{fig:fig1}(d)).
Notably, the chiral states also exists for finite range interactions if all particles are located within a small circle and have sufficiently small self-propulsion speed $v_0$ ~\cite{supp}.

To render this complex finite-$N$ behavior consistent with the mean-field predictions ($N\to\infty$) we analyzed the finite-size behavior of the 
probability distribution of the stationary angular frequency
$\Omega=\langle\dot\psi(t)\rangle$, where $\langle\ldots\rangle$ denotes a time average.
Fig.~\ref{fig:fig2}(j) shows that the support of $P(\Omega)$ shrinks with
increasing systems size $N$, which implies that any rotational 
motion of the order parameter becomes on average slower with 
increasing system size and ceases in the limit $N\to\infty$.
We note that even for large but finite system sizes still slow rotations exist, which renders a precise extrapolation of observables to infinite
system size by integrating (\ref{eq2}) computationally unfeasible. Nevertheless, the proper
$N\to\infty$ limit is given by the solution of eq.(\ref{eq3}).


\section{Finite interaction range}
{\bf Phase diagram.}
In the case of a finite interaction range, $R$, we integrate (\ref{eq1}, \ref{eq2}) using Euler method with time step $dt=0.05$, propulsion speed $v_0=0.5$. We set $R=1$ and discuss here only 
uncorrelated randomness $\lambda=0$. For $\alpha=1$
flocking already occurs for small alignment bias $J_0\ll J(=1)$,
which decreases with increasing density, $\rho=N/L^2$, as shown in Fig.~\ref{fig:fig3}(a), reminiscent of
the decrease of $J_{0,c}(N)$ with $N$ for the infinite range case.
In the flocking phase, except for small $\rho$, the system eventually evolves into a single dense cluster in which all particles are tightly packed and move coherently(inset of Fig.~\ref{fig:fig3}(a)).
This single-cluster phase emerges via nucleation: once a denser
cluster forms, the contribution of the alignment 
bias to the interactions dominates over the random contribution 
due to the $\alpha=1$ scaling, such that the cluster  eventually absorbs all particles(Fig.~\ref{fig:fig3}(c)).

For $\alpha=1/2$, the critical alignment bias, $J_{0,c}$, also 
decreases with increasing density, $\rho$, as shown in (Fig.~\ref{fig:fig3}(d)).
In both cases, for $\alpha=1/2$ and $\alpha=1$, the transition from the disordered phase to the flocking phase is discontinuous(Fig.~\ref{fig:fig3}(e)), reminiscent of the Vicsek model~\cite{Gregoire2004onset,Chate2008collective}.
In contrast to the latter, when crossing the transition line in Fig.~\ref{fig:fig3}(d) for instance by increasing the density, $\rho$, spontaneously global polar order emerges without an intermediate band formation.
As we increases $J_0$ further, we also find a transition to a single-cluster phase, see Fig. S4 in the Supplement \cite{supp}.
Here the single-cluster phase arises due to coarsening, where small clusters move ballistically and merge into a larger cluster upon collision as in the conventional Vicsek model at very low noise~\cite{Barberis2018emergence}.

{\bf Clique formation.}
Both disordered and ordered phase in the low density regime are characterized by clique formation, where transient clusters of coherently moving particles emerge and disintegrate dynamically even in absence of a ferromagnetic bias $J_0=0$.
To analyze these cliques, we construct interaction networks based on distance between particles at successive simulation times and track the evolution of their connected components\cite{Buchin2015trajectory}(Fig.~\ref{fig:fig4}(a)).
Among the connected components that persist over time, we define a 'clique' $\mathcal C(n,\tau)$ as a set of particles in a connected component with size $n$ and lifetime $\tau$, restricted to those with $n\ge5$ and $\tau\ge10$(Fig.~\ref{fig:fig4}(c)).
Fig.~\ref{fig:fig4}(b) represents an example time series for a specific particle, illustrating that only a small subset of its interacting neighbors actually form cliques with it.
Fig.~\ref{fig:fig4}(d) shows $P(n,\tau)$, the probability that a particle belongs to $\mathcal C(n,\tau)$,
averaged over time and disorder. 
Both $P(n)$ and $P(\tau)$ exhibit exponential tail(Fig.~\ref{fig:fig4}(e),(g)) and the characteristic clique size $n^\ast$ and lifetime $\tau^\ast$ are extracted from exponential fits(Fig.~\ref{fig:fig4}(f),(h)). Notably, here the life-time $\tau$
of medium- and small-size cliques is actually sufficiently long
for them to traverse half of the system.

The formation of a clique is driven by the self-assembly of particles with predominantly mutual aligning interactions 
$\langle J_+(n,\tau)\rangle>0$, where $J_+(n,\tau)$ is an 
average of $2J_0+J_{ij}+J_{ji}$ over the members $i,j$ of 
a clique, as shown in the inset of Fig.~\ref{fig:fig4}(d).
Cliques disintegrate when they collide with other strongly interacting particles or cliques~\cite{supp}. 
When compared with clusters emerging in the disordered phase of the conventional Vicsek model with homogeneous alignment interactions one notes that in the latter clusters are typically larger but have shorter life-times.

%
%

\smallskip

\section{Discussion}

We have shown that flocking can occur in an ensemble of self-propelled particles with a random mixture of non-reciprocal aligning and anti-aligning. Within the ordered phase, for sufficiently strong alignment bias, spurious chiral states emerge in finite systems with infinite-range interaction, separated from conventional flocking by exceptional points or Hopf bifurcations. Those states disappear in the infinite system size limit, where purely relaxational dynamics towards global flocking persists. For finite interaction range coherently moving cliques emerge, in which a small group of particles self-organize transiently, even in the absence of a global alignment bias in the interactions: equipped with random and fully non-reciprocal ($\lambda=0$) interactions $J_{ij}$, particles with $J_{ij}$ and $J_{ji}$ positive (aligning) find each other and move for some time together until they collide with other groups. It should be noted that this clique formation is different from conventional cluster formation, in which particles can join and leave clusters: a clique is defined as a specific group of particles that stays together until the clique dissolves \cite{remark}.

In a forthcoming publication~\cite{tobepublished} we will discuss the robustness of these phenomena in the presence of noise or the introduction of the Vicsek update rules for directional changes. Our preliminary results show that the ($\lambda,J_0$, non-reciprocity / alignment bias) phase diagrams that we reported are robust with respect to the introduction of weak noise as well as Vicsek update rules. Also, the chiral states of finite systems in the ordered phase of the infinite range limit are stable against weak noise as well as the clique formation for finite interaction range. Stronger noise will, however, have an impact. Also, the variation of the degree of non-reciprocity bears interesting aspects as an enhanced tendency to order for $\lambda\to1$ and an emerging spin glass phase at $\lambda=1$.
 
Our results may serve as a starting point for the study of multi-species flocking models with non-random but complex non-reciprocal inter-species interactions in the following sense: As non-reciprocal interactions are used to generate temporal sequences of patterns in neural network models~\cite{Sompolinsky1986temporal} or for dynamical control of self-assembled immobile structures and transitions between~\cite{Osat2022nonreciprocal} we expect that programmable non-reciprocal inter-actions in flocking models as the one analyzed here could be a tool to create mobile ”shape-shifters” made of self- propelled particles that arrange in a pre-determined temporal sequence of patterns with varying shapes. It would be worthwhile to examine this vision in future work.




\bibliography{apssamp}

\input{SM}

\end{document}

%% file: SM.tex
\renewcommand{\theequation}{S\arabic{equation}}
\renewcommand{\thefigure}{S\arabic{figure}}
\setcounter{equation}{0}
\setcounter{figure}{0}
\setcounter{secnumdepth}{2}
\onecolumngrid
\newpage
\begin{center}
{\bf\large Supplemental material:\\
Flocking with random non-reciprocal interactions}\\
\vspace{4mm}
\setcounter{page}{1}

Jiwon Choi$^1$, Jae Dong Noh$^2$, Heiko Rieger$^1$\\
\vspace{2mm}
{\it $^1$ Department of Physics \& Center for Biophysics,\\
Saarland University, Campus E2 6, 66123 Saarbrücken, Germany\\
$^2$ Department of Physics, University of Seoul, Seoul, 02504, Korea}\\
\end{center}

\section{Finite size scaling at the critical point $J_{0,c}$}.


In the infinite-range limit for $\alpha=1$, relative strength of non-reciprocal alignment over reciprocal alignment bias scales as $N^{-1/2}$.
Thus, phase behaviors of the system
with $\alpha=1/2$ at alignment bias $J_0$ are equivalent to those of the system with $\alpha=1$ at alignment bias 
$J_0\sqrt{N}$.
We confirm this equivalence by performing finite-size scaling(FSS) analysis.

\begin{figure}[h]
    \includegraphics[width=\linewidth]{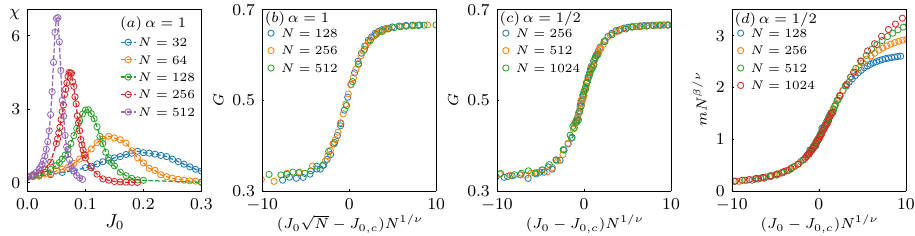}
    \caption{\label{fig:sm_fig1}
        Finite-size scaling analysis for $\lambda=0$ case.
        (a) Order parameter susceptibility $\chi$ for $\alpha=1$.
        (b-c) Scaling collapse of Binder cumulant $G$ for (b) $\alpha=1$ with rescaled $J_0\to J_0\sqrt N$, and (c) $\alpha=1/2$.
        $J_{0,c}=1.1$ is the DMFT critical point and $\nu=5/2$ is estimated.
        (d) Scaling collapse of magnetization $m$ for $\alpha=1/2$ with $\beta=1/2$.
    }
\end{figure}

For $\alpha=1$, to locate $N$-dependent critical point $J_{0,c}(N)$, we compute the order parameter susceptibility, 
\begin{equation}
    \chi(J_0,N) = N\left( [\langle m^2\rangle] - [\langle m\rangle^2] \right)
\end{equation}
where $\langle\cdots \rangle$ denotes a time average, $[\cdots]$ denotes an average over random interaction matrices, and $m$ represents the amplitude of order parameter $m(t)e^{i\psi(t)} = N^{-1}\sum_j e^{i\theta_j(t)}$.
We define $J_{0,c}(N)$ for $\alpha=1$ as the values of $J_0$ at which the order parameter susceptibility reaches its maximum(Fig.\ref{fig:sm_fig1}(a)), and we use this definition in Fig.~1(a) in the main text.

To check consistency with the DMFT calculation and estimate critical exponents, we use the following scaling relations
\begin{equation}
    G = \mathcal G[(J_0-J_{0,c})N^{1/\nu}]
    \label{eq:sm/G_scaling}
\end{equation}
\begin{equation}
    m = N^{-\beta/\nu}\mathcal M[(J_0-J_{0,c})N^{1/\nu}]
    \label{eq:sm/m_scaling}
\end{equation}
where $J_{0,c}=1.1$ is the critical point from DMFT(see Fig.1 in the main text), and $G$ denotes Binder cumulant defined as
\begin{equation}
    G = 1-\left[\frac{\langle m^4\rangle}{3\langle m^2\rangle^2}\right]
\end{equation}
Fig.~\ref{fig:sm_fig1}(b,c) show scaling collapse of Binder cumulant for $\alpha=1$(with rescaled $J_0\to J_0\sqrt N$) and $\alpha=1/2$ using \eqref{eq:sm/G_scaling}.
In both cases, $J_0=1.1$ and $\nu=5/2$ produce good scaling collapses.
The order parameter exponent $\beta=1/2$ is also determined in Fig.~\ref{fig:sm_fig1}(d) using
\eqref{eq:sm/m_scaling}.

For $\lambda\neq0$ cases, the transition line in Fig.1(c) in the main text is obtained using the same analysis.

\section{Chiral states in finite systems (infinite range, $\lambda=0$)}

\begin{figure}
    \includegraphics[width=\linewidth]{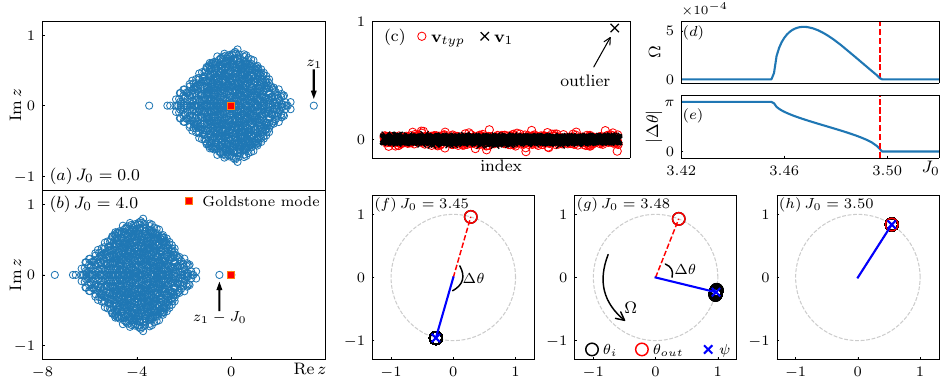}
    \caption{\label{fig:sm_fig2}
        Eigenvalue spectrum of $\mathbf J^\ast$ with (a) $J_0=0$ and (b) $J_0=4.0$.
        The red square depicts the Goldstone mode $\mathbf v_0=N^{-1/2}(1,\cdots,1)$ and $z_1$ denotes the largest eigenvalue of $\mathbf J^\ast$.
        (c) Components of a typical bulk eigenvector $\mathbf v_{\text{typ}}$(red circles) and the leading eigenvector $\mathbf v_1$ associated with $z_1$ at $J_0=0$(black pluses). 
        $\mathbf v_1$ exhibits highly localized structure, and we refer to its largest component as the 'outlier component'.
        This outlier component is depicted as a red circle in (f-h), where it drives global chiral motion.
        (d) Steady-state angular frequency $\Omega\equiv \langle\dot\psi(t)\rangle$.
        (e) Angle difference between the outlier component and the order parameter $\Delta\theta \equiv \theta_{\text{out}} - \psi$.
        (f-h) Steady-state configurations $\{\theta_i\}$ at different $J_0$ values: (f) $J_0=3.45$, (g) $J_0=3.48$, and (h) $J_0=3.50$.
        See also `movie1'.
        The red circle depicts the outlier component $\theta_{\text{out}}(t)$, corresponding to maximum component of $\mathbf v_1$(see inset of (a)).
    }
\end{figure}
In this section, we show that the rotational motion at $J_0=z_1$ originates from the critical exceptional point(CEP). 
We mainly focus on a single realization of interaction matrix, shown in Fig.~2(e) in the main text, but the qualitative behaviors remain robust for other realizations.

In mean-field limit with $\alpha=1/2$, the equation of motion reduces to
\begin{equation}
    \dot\theta_i(t) =
    -\sum_j \left(\frac{J_{ij}}{\sqrt{N}}
    +\frac{J_0}{N}\right)\sin(\theta_i(t)-\theta_j(t))
    \label{sm_eq1}
\end{equation}
When $J_0\gg J_{0,c}$, phase differences between oscillators become negligible, thus the system remains perfectly syncronized state $\theta_i(t) = \psi(t) + \delta_i(t)$, where $\psi(t)$ denotes the phase of magnetization.
In this limit, we can approximate (\ref{sm_eq1}) as
\begin{equation}
    \dot{\delta}_i \approx -\sum_j\left( \frac{J_{ij}}{\sqrt N} + \frac{J_0}N \right)\left(\delta_i - \delta_j\right) \quad\Rightarrow\quad \dot{\vec\delta} = \mathbf J^\ast \vec\delta
\end{equation}
where the components of linearized interaction matrix $\mathbf J^\ast$ are given by
$[\mathbf J^\ast]_{ij} = J_{ij}/\sqrt N + J_0/N$ for $i\neq j$ and $[\mathbf J^\ast]_{ii} = \sum_{j\neq i}[\mathbf J^\ast]_{ij}$.
The diagonal component forces the zero-row-sum condition $\sum_j[\mathbf J^\ast]_{ij} = 0$, which originates from the rotational invariance of (\ref{sm_eq1}).
As a consequence, the Goldstone mode $\mathbf v_0=N^{-1/2}(1,\cdots,1)$ associated with global rotational invariance exists with zero eigenvalue $z_0=0$, marked in Fig.~\ref{fig:sm_fig2}(a,b).

Although the eigenvalue spectrum of the interaction matrix $[\mathbf J]_{ij}=J_{ij}/\sqrt N$ exhibits the circular law~\cite{Sommers1988spectrum}, rotational invariance significantly deforms the eigenvalue spectrum.
In addition, the presence of ferromagnetic bias $J_0$ acts as a rank-1 perturbation on $\mathbf J^\ast$, which uniformly shifts the whole spectrum by $-J_0$ along the $\text{Re}\,z$ axis, except for the Goldstone mode(Fig.~\ref{fig:sm_fig2}(a,b)).
As a result, when $J_0>z_1$, where $z_1$ denotes the largest eigenvalue, all eigenvalues have a negative real part and only the Goldstone mode remains at $z=0$.
As $J_0(>z_1)$ decreases, $z_1$ approaches $z=0$ and eventually collides with the Goldstone mode at $J_0=z_1$.
And the eigenvector $\mathbf v_1$ associated with $z_1$ coalesces with the Goldstone mode $\mathbf v_0$ at this point, indicating the CEP(see Fig.2 (f,g) in the main text)~\cite{Hanai2020critical,Suchanek2023time,Zelle2024universal}.

As soon as $J_0$ crosses the CEP, the system exhibits spontaneous chiral motion, where all oscillators rotate with a constant angular velocity $\Omega$(Fig.~\ref{fig:sm_fig2}(g) and `movie1').
The direction of chiral motion depends on the initial condition, and we generally observe counter-clockwise and clockwise rotation with an equal probability.
When the chiral motion takes a place, the steady-state configuration $\{\theta_i\}$ has a special structure:
While all other oscillators remain close each other, an 'outlier' $\theta_{\text{out}}$(the maximum component of $\mathbf v_1$ in Fig.~\ref{fig:sm_fig2}(c)) maintains non-zero angle difference with them, which is described by $\Delta\theta = \theta_{\text{out}}-\psi$(Fig.~\ref{fig:sm_fig2}(f-h)).
The angle difference increases as $J_0$ decreases(Fig.\ref{fig:sm_fig2}(e)) and reaches $\pi$.
The steady-state angular frequency approaches to zero as $\Delta\theta$ approaches $\pi$(Fig.\ref{fig:sm_fig2}(e)), and static flocking phase is restored where only 'outlier' points opposite direction with others(Fig.\ref{fig:sm_fig2}(f)).

The second instability appears as the second largest eigenvalue hits zero.
Its nature depends on disorder realization: When second largest eigenvalue is real(or complex with its conjugate), then it exhibits CEP(oscillatory instability).
Observed complex behaviors in the main text appear in bulk spectrum region, where multiple instabilities contribute.



\section{Chiral states with finite interaction range}

In this section, we discuss the possibility of spontaneous chiral motion with a finite interaction range.
As in the main text, we set the interaction range to $R=1$.
We find that a spontaneous chiral state can emerge in finite interaction range, but only under specific initial conditions: All particles are located in a small region(smaller than $R$) and the propulsion speed $v_0$ should be sufficiently small.
The initial orientations can be randomly assigned.

\begin{figure}
    \includegraphics[width=\linewidth]{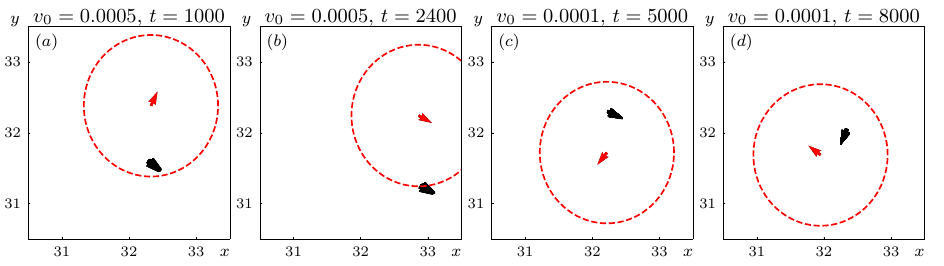}
    \caption{\label{fig:sm_fig3}
        Snapshots from 'movie3' for different self-propulsion speeds: (a,b) $v_0=0.0005$ and (c,d) $v_0=0.0001$.
        The red arrow and circle depict $\theta_{\text{out}}$ and its interaction range.
        For larger $v_0$(a,b), $\theta_{\text{out}}$ partially loses its interacting neighbors as it moves, and whole system restores perfect alignment.
        In contrast, for small $v_0$(c,d), all particles remains within their interaction range and chiral motion is stable.
    }
\end{figure}

Fig.~\ref{fig:sm_fig3}(see also 'movie3') shows such a case at $J_0=3.47$ using the same interaction matrix as Fig.~\ref{fig:sm_fig2} for different $v_0$.
Starting from the given initial conditions, the system gradually develops a spontaneous chiral motion with the same angular frequency $\Omega$, provided that all inter-particle distances stay within the interaction range $R$.
Since each particle follows a circular motion of radius $r = v_0/\Omega$, the condition that this chiral state remains stable is given by $v_0\lesssim \Omega$.
If $v_0$ exceeds the threshold, $\theta_{\text{out}}$ loses some interaction neighbors(Fig.~\ref{fig:sm_fig3}(a,b)), and the coherent chiral motion breaks down.
In contrast, for the case where $v_0$ is sufficiently small, the chiral motion remains stable(Fig.~\ref{fig:sm_fig3}(c,d)).

\section{Single-cluster phase for $\alpha=1/2$ with finite interaction range}
\begin{figure}
    \includegraphics[width=0.7\linewidth]{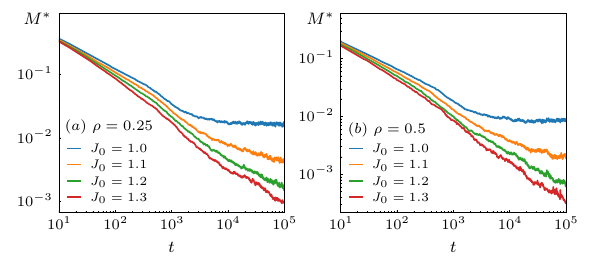}
    \caption{\label{fig:sm_fig4}
        Normalized number of cluster $M^\ast$ for (a) $\rho=0.25$ and (b) $\rho=0.5$ with $L_x=L_y=128$, $\lambda=0$, $v_0=0.5$, and $R=1$.
        Each line indicates average over 25 different disorder realizations starting from the random initial condition.
    }
\end{figure}

In this section, we discuss emergence of single-cluster phase for $\alpha=1/2$ with finite interaction range $R=1$.
To quantifying emergence of single-cluster state at $\alpha=1/2$, we calculate the normalized number of clusters $M^\ast = M/N$, where $M$ is the number of clusters and $N=L_xL_y\rho$ is the total number of particles~\cite{Barberis2018emergence}.
Fig.~\ref{fig:sm_fig4} represents $M^\ast$ for $\rho=0.25,0.5$ with various $J_0$ values.
In both cases, for $J_0=1.2$, $M^\ast$ exhibits power-law decay without saturation, which indicates that the system approach to the single-cluster phase as in the conventional Vicsek model at low noise~\cite{Barberis2018emergence}.
Thus, in $J_0\ge1.2$ regime, the single-cluster phase emerges via active coarsening: small clusters move ballistically and merge upon collision.

\section{Description of supplemental movies}
The interaction matrix used in movies (1-3) is identical to that in Fig.~2 of the main text.
Movies are also available at \href{https://doi.org/10.5281/zenodo.15742179}{https://doi.org/10.5281/zenodo.15742179}.
\begin{itemize}
    \item movie 1: Emergence of spontaneous chiral motion near the critical exceptional point(CEP).
    Parameters: $N=1024$ and $\lambda=0$.
    \item movie 2: Various phase behaviors in flocking phase: (a) flocking with relaxational dynamics, (b) vibrating oscillation, (c) oscillation, and (d) rotation.
    Parameters: $N=1024$ and $\lambda=0$.
    \item movie 3: Chiral state in finite-range interactions.
    The system starts from an initial condition where particles are confined within a small region and have random initial orientations.
    Parameters: $N=1024$, $\lambda=0$, and $J_0=3.47$.
    \item movie 4: Clique formation in steady-state at $J_0=0$.
    The color represents the size of each clique, and black arrows indicates long-living pair(or triplet) with lifetime $\tau>150$.
    Parameters: $L_x=L_y=128$, $\lambda=0$, $\rho=0.25$, $R=1$, and $J_0=0$.
\end{itemize}



%